\documentclass{aa}
\usepackage{graphicx}
\usepackage{natbib,twoopt}
\usepackage[breaklinks=true]{hyperref} 
\usepackage[varg]{txfonts}
 \bibpunct{(}{)}{;}{a}{}{,}    
 \newcommandtwoopt{\citeads}[3][][]{\href{http://adsabs.harvard.edu/abs/#3}%
                                        {\citealp[#1][#2]{#3}}}
 \newcommandtwoopt{\citepads}[3][][]{\href{http://adsabs.harvard.edu/abs/#3}%
                                        {\citep[#1][#2]{#3}}}
 \newcommandtwoopt{\citetads}[3][][]{\href{http://adsabs.harvard.edu/abs/#3}%
                                        {\citet[#1][#2]{#3}}}
 \newcommandtwoopt{\citealtads}[3][][]{\href{http://adsabs.harvard.edu/abs/#3}%
                                        {\citealt[#1][#2]{#3}}}
 \newcommandtwoopt{\citeyearads}[3][][]%
   {\href{http://adsabs.harvard.edu/abs/#3}{\citeyear[#1][#2]{#3}}}
\begin{document}

\title{Large-scale magnetic fields in Bok globules}

\titlerunning{Large-scale magnetic fields in Bok globules}

\author{G. Bertrang\inst{\ref{inst1},}\thanks{gbertrang@astrophysik.uni-kiel.de}
        \and S. Wolf\inst{\ref{inst1}}
        \and H. S. Das\inst{\ref{inst2}}
       }

\institute{University of Kiel, Institute of Theoretical Physics and Astrophysics, Leibnizstr. 15, 24118 Kiel, Germany\label{inst1}
           \and Assam University, Department of Physics, Silchar – 788 011, India \label{inst2}
          }

\authorrunning{Bertrang et al.}

\date{Received 20/11/2013 / Accepted 26/03/2014}

\abstract
{The role of magnetic fields in the star formation process is a contentious matter of debate. In particular, no clear observational proof exists of a
general influence by magnetic fields during the initial collapse of molecular clouds.}
{Our aim is to examine magnetic fields and their influence on a wide range of spatial scales in low-mass star-forming regions.}
{We trace the large-scale magnetic field structure on scales of $10^3-10^5$~AU in the local environment of Bok globules through optical and
near-infrared polarimetry and combine these measurements with existing submillimeter measurements, thereby characterizing the small-scale magnetic field
structure on
scales of $10^2-10^3$~AU.}
{For the first time, we present polarimetric observations in the optical and near-infrared\thanks{Based on observations made with ESO telescopes at the La
Silla Paranal Observatory under programme IDs 089.C-0846(A) and 090.C-0785(A).} of the three Bok globules B335, CB68, and CB54, combined with
archival observations in the submillimeter and the optical. We find a significant polarization signal $\left(P\gtrsim2~\%,\ P/\sigma_{\rm{P}}>3\right)$ in the
optical and near-infrared for all three globules. Additionally, we  detect a connection between the structure on scales of $10^2-10^3$~AU to
$10^3-10^4$~AU for both B335 and CB68. Furthermore, for CB54, we trace ordered polarization vectors on scales of $\sim10^5$~AU. We
determine a magnetic field orientation that is aligned with the CO~outflow in the case of CB54, but nearly perpendicular to the CO~outflow for CB68. For B335 we
find a change in the magnetic field oriented toward the outflow direction, from the inner core to the outer
regions. }
{We find strongly aligned polarization vectors that indicate dominant magnetic fields on a wide range of spatial scales.}


\maketitle

\section{Introduction}

Magnetic fields are observed in a wide range of astronomical scales: from the entire galaxy and giant molecular clouds
\citep{1996ARA&A..34..155B, 2001SSRv...99..243B}, to smaller molecular clouds with low-mass star formation and protostellar disks \citep[see, e.g.,
][]{2008Ap&SS.313...87G, 2001ApJ...561..871H, 2004Ap&SS.292..239W}, to circumstellar disks where they drive outflows, jets, and the accretion onto the central
star \citep[see, e.g.,][]{2007prpl.conf..277P, 2009ApJ...691L..49S, 1998ApJ...492..323G}. Magnetic fields have been discussed as playing an important role in
star formation \citep[see, e.g.,][]{2009MmSAI..80...54G}. In general, they can influence the contraction timescale, the gas-dust coupling, and the shape of
cloud fragments, and they host jets and outflows \citep[see, e.g.,][]{2007ARA&A..45..565M}.\\
In the dusty envelopes around young stellar objects, polarization of background starlight due to dichroic extinction and thermal emission by non-spherical dust
grains is the most important signature of magnetic fields \citep[see, e.g.,][]{2000prpl.conf..247W, 2008Ap&SS.313...87G,  1999AAS...194.4714C}. The dust grains
become partially aligned to the magnetic field, with their long axes perpendicular to the magnetic field lines \citep[see, e.g.,][]{2007JQSRT.106..225L}.
The thermal emission of the dust grains thus becomes polarized, with a polarization direction perpendicular to the magnetic field, projected onto the plane
of sky. Additionally, the light of background stars that shines through the star-forming region becomes polarized by dichroic absorption by the dust grains.
Here, the resulting polarization maps directly trace the magnetic field lines, projected onto the plane of sky \citep[see,
e.g.,][]{2000prpl.conf..247W}. Such polarization observations allow an assessment of the relative importance of uniform and tangled magnetic fields: A high
level of polarization, uniform in direction, indicates a well-ordered field that is not significantly tangled on scales smaller than the beam size, i.e., a
magnetic field that plays an important role in the evolution of the local density structure during the star formation.
\\
A very good environment for studying the role of magnetic fields is given in low-mass star-forming regions, so-called Bok globules. These objects are less
affected by large-scale turbulences and other nearby star-forming events. Bok globules are small in diameter ($0.1-2$~pc), simply structured, and are relatively
isolated molecular clouds with masses of $2-100$~M$_\sun$ \citep{1991ApJS...75..877C, 1977PASP...89..597B, 1985prpl.conf..104L}.\\
Given the sensitivity of millimeter/submillimeter (sub-mm) telescopes that allow for polarimetric observations, the thermal emission of the dust grains,
observable in the sub-mm, traces the densest, central part of a Bok globule. The less dense, outer parts are traced with polarized observations of
background starlight that is dichroicly absorbed and observable in the near-infrared (near-IR)~/ optical. Thus, multiwavelength observations that
combine sub-mm, near-IR, and optical polarization observations reveal the magnetic field geometry from the smallest to the largest scales of the Bok globule.
So far, there have only been about two dozen sub-mm polarization observations of Bok globules, while only about half a dozen of these show ordered magnetic
field structures \citep{2009ApJS..182..143M, 2010ApJS..186..406D}. The majority of the sub-mm polarization observations show tangled field patterns, an
indicator for negligible magnetic fields. Contrary to that, the near-IR and optical observations of \citet{2011AJ....142...33A}, \citet{2013ApJ...769L..15S},
and \citet{2000A&AS..141..175S} revealed ordered field structures in the less dense, outer parts, indicating dominant magnetic fields in these parts of low-mass
star-forming regions. Motivated by this, we performed a polarization study of Bok globules with two morphologically different globule types: B335 and
CB68 are simply structured, small globules, whereas CB54 is a more complexly structured, larger Bok globule. In this work, we present polarization observations
in the optical and near-IR of the three Bok globules B335, CB68, and CB54, combined with archival sub-mm and optical
observations, for the first time.\\
We start with a description of the sources and the selection criteria in Section~\ref{sec:description}. In Section~\ref{sec:obs} we describe the observations,
and in Section~\ref{sec:datared} the data reduction. In Section~\ref{sec:polmap} we analyze the polarization maps, and in Sects.~\ref{sec:Bfield}
and~\ref{sec:CO} we discuss the magnetic field and the correlation between the magnetic field structure and the CO~outflows. In Section~\ref{sec:summary} we
discuss the observability of the gap regions between the sub-mm and near-IR observations that we find and summarize our results.

\section{Description of the sources}\label{sec:description}
Our aim is to observe the polarization structure and, by this, the magnetic field structure on a wide range of scales with as few external influences as
possible, e.g., nearby star-forming activity or reflection nebulae. This restricts us to isolated and compact Bok globules with available archival sub-mm
polarization maps that show regular magnetic field structures. Additionally, the globules ought to have a high number of potential background stars in the
particular wavelength range and to be observable from sites with access to optical or near-IR polarimeters. We applied these selection criteria to
the
legacy catalogs of the $350~\mu$m~polarimeter, Hertz, at the Caltech Submillimeter Observatory \citep[CSO,][]{2010ApJS..186..406D} and the
$850~\mu$m~polarimeter for SCUBA, SCUPOL, at the James Clerk Maxwell Telescope \citep[JCMT,][]{2009ApJS..182..143M} and find that these criteria are satisfied
only by the three Bok globules B335, CB68, and CB54.\\
{\bf B335} is a Bok globule at a distance of only $\sim100$~pc \citep{2009A&A...498..455O}, and it accommodates one of the best-studied low-mass protostellar
cores \citep[see, e.g.,][]{2003ApJ...596..383H, 2005ApJ...626..919E}. Its protostar drives a collimated bipolar outflow with a dynamical age of $\sim3\times
10^4$~yr \citep[see, e.g.,][]{1993ApJ...414L..29C}. The dense core in B335 is generally recognized as the best protostellar collapse candidate, and the emission
from different molecular lines has been  successfully modeled in terms of an inside-out collapse \citep{1977ApJ...214..488S} with an infall age of $\sim10^5$~yr
and a current protostar mass of $\sim0.4$~M$_{\sun}$~ \citep[see, e.g.,][]{1995ApJ...448..742C}. A total envelope mass of $\sim4$~M$_{\sun}$\ within a radius of
$1.5\times10^4$~AU was derived from sub-mm observations by \citet{2003ApJ...592..233W}.\\
{\bf CB68} is a globule in a distance of only $\sim160$pc \citep{1997A&A...326..329L}. It is remarkable for the very high percentage of polarization near
850~$\mu$m \citep{2000ApJ...542..352V}. The small core of CB68 with a mass of $\sim0.9$~M$_\sun$ and a diameter of $0.03$~pc is just below the critical size
scale of $0.05$~pc, identified as the diameter where there is a break in star formation processes and below which there is a coherent core
\citep{2000ApJ...542..352V}.
Roughly perpendicular to its bipolar CO~outflow, there is an elongated structure seen in both dense gas and dense dust, indicating a pseudodisk
\citep{2003ApJ...588..910V}. Sub-mm polarization data obtained with SCUBA/JCMT implies an hourglass-type magnetic field structure \citep{2007AJ....134..628V}.\\
{\bf CB54} is a large Bok globule, associated with the molecular cloud BWW~4 in a distance of $\sim1.1$~kpc \citep{1993A&A...275...67B}. The globule contains a
massive dense core of $M_{\rm{H}}\sim100$~M$_\sun$ that drives a bipolar molecular outflow \citep{1997A&A...326..329L, 1994ApJS...92..145Y}. At near-IR
wavelengths, a small, deeply embedded young stellar cluster becomes visible \citep{2004Ap&SS.292..239W}. CB54 shows a randomly oriented polarization
pattern in the sub-mm \citep[][see Fig.~\ref{fig:CB54_polmap}~b]{2001ApJ...561..871H}. However, in the optical the polarization pattern appears more homogenous
\citep[][see Fig.~\ref{fig:CB54_polmap}~c]{2000A&AS..141..175S}.

\section{Observations}\label{sec:obs}
The observations were performed in the near-IR with ISAAC/VLT and SOFI/NTT and in the optical with IFOSC/IGO (see Table~\ref{table:observations}).

\begin{table}
\caption{Summary of the observations.}         
\label{table:observations}     
\centering                      
\begin{tabular}{l l c l}      
\hline\hline                
Object & Instrument & Filter & Date\\  
\hline                                    
B335 & ISAAC/VLT & Js & 2012 Mar-May\\     
CB68 & ISAAC/VLT & Js & 2012 Mar-May\\
     & IFOSC/IGO & R  & 2012 Mar 21-22\\
CB54 & SOFI/NTT  & Js & 2013 Jan 24-26\\
\hline                                  
\end{tabular}
\end{table}
\begin{table*}[!ht]
\caption{Polarization standard stars.}             
\label{table:standards_summary}      
\centering          
\begin{tabular}{l l c c c c c c c }
\hline\hline       
Instrument & Object & $\alpha_{2000}$ & $\delta_{2000}$ & Type & P & $\gamma$ & Filter & Ref.\\
& & (hh:mm:ss.ss) & (dd:mm:ss.ss) & & (\%) &  $\left(^\circ\right)$ & & \\
\hline                    
ISAAC/VLT & EGGR118      & 16:17:55.26 & -15:35:51.93 & unpolarized & $<1.79$         &             & J & 1 \\  
	 	  & GJ1178       & 13:47:24.36 & +10:21:37.90 & unpolarized & $<1.08$         &             & J & 1 \\
SOFI/NTT  & CMa R1 No.24 & 07:04:47.36 & -10:56:17.44 & polarized   & $2.1\pm0.05$    & $86\pm1$    & J & 2 \\
	      & HD64299      & 07:52:25.51 & -23:17:46.78 & unpolarized & $0.151\pm0.032$ &             & B & 3 \\
	      & WD0310-688   & 03:10:31.02 & -68:36:03.39 & unpolarized & $0.051\pm0.09$  &             & V & 4 \\
IFOSC/IGO & HD251204     & 06:05:05.67 & +23:23:38.54 & polarized   & $4.04\pm0.066$  & 147         & V & 3 \\
	      & GD319        & 12:50:05.00 & +55:06:00.00 & unpolarized & $0.045\pm0.047$ &             & B & 3 \\
	      & HD65583      & 08:00:32.12 & +29:12:44.40 & unpolarized & $0.013\pm0.02$  & $144.7\pm30$ & B & 5 \\
\hline                  
\end{tabular}

\tablebib{(1)~\citet{2008MNRAS.387..713R}; (2) \citet{1992ApJ...386..562W}; (3) \citet{1990AJ.....99.1243T}; (4) \citet{2007ASPC..364..503F}; (5)
\citet{1981AJ.....86.1518C}.}
\end{table*}

\subsection{Near-IR observations}
The near-IR observations were carried out in March-May 2012  and January 2013 at ESOs Very Large Telescope (VLT) and the New Technology Telescope (NTT). We
observed with the instruments Infrared Spectrometer And Array Camera (ISAAC) and Son OF ISAAC (SOFI), which are mounted at the Nasmyth~A foci of the
$8~$m VLT (UT3), respectively of the $3.58$~m NTT, both equipped with a $1024\times 1024$ Hawaii Rockwell array optimized for wavelengths of $1-2.5~\mu$m.\\
In both ISAAC/VLT and SOFI/NTT a single Wollaston prism is used for polarization observations. In this observing mode, the polarized flux is measured
simultaneously at two different angles that differ by $90\degr$. To derive the linear polarization degree and orientation of an object, two observations
must be performed at each pointing of the telescope with different orientations of the Wollaston prism, typically $0\degr$ and $45\degr$. For ISAAC/VLT and
SOFI/NTT, this is realized by a rotation of the complete instrument. To avoid overlapping between different polarization images an aperture mask of
three alternating opaque and transmitting strips of about $20''\times150''$ for ISAAC/VLT and $40''\times300''$ for SOFI/NTT is used.\\
We carried out Js-band polarization observations of five fields of B335 and CB68 with ISAAC/VLT and of four fields of CB54 with SOFI/NTT.

\subsection{Optical observations}
The optical observations in R~band were carried out in March 2012 at the $2~$m~IUCAA Girawali Observatory (IGO, India). The IUCAA Faint Object Spectrograph
\& Camera (IFOSC), attached at IGOs Cassegrain focus, is equipped with an EEV~$2$K~$\times~2$K CCD camera optimized for a wavelength range
of \mbox{$350-850$~nm} \citep[for details, see][]{2002BASI...30..785}.\\
IFOSC/IGOs polarimetry mode makes use of a single Wollaston prism combined with a half-wave plate. The linear polarization degree and orientation is derived
through two observations with two different half-wave plate orientations. The field of view for the polarimetry mode is about $4'\times4'$. IFOSC/IGO does not
use an aperture mask like ISAAC/VLT or SOFI/NTT. The unambiguous assignment of the observed point sources to the corresponding direction of
polarization is done as part of the data reduction procedure.\\
We performed polarization observations of four fields of CB68 with IFOSC/IGO.

\section{Data reduction}\label{sec:datared}

The data reduction was performed with pipelines specialized for each instrument \citep[for details of
polarimetric data reduction and calibration see, e.g.,][]{HowToPol}. These pipelines are used for bias correction, flat-fielding, and instrumental
polarization removal. The Stokes parameters $I, Q,$ and $U$ are computed through aperture photometry. During the observations with ISAAC/VLT, we observed at
orientations of the Wollaston prism of $0\degr, 45\degr$. During the SOFI/NTT and IFOSC/IGO observations, we observed respectively with the Wollaston prism and 
the half-plate at four different orientations of $0\degr, 22.5\degr,45\degr, 67.5\degr$. Unpolarized and polarized standard stars have been observed to
determine the instrumental polarization. Table~\ref{table:standards_summary} summarizes
the general information for these stars.\\
\begin{table}[!htpb]
\caption{Observational results for the polarized standard star of CB54.}
\label{table:standards_observations}     
\centering                      
\begin{tabular}{l l c c c c}
\hline\hline                
Object & Obs. date & $P$  & $\sigma_{\rm{P}}$ & $\gamma$              & $\sigma_{\rm{\gamma}}$ \\
       &           & (\%) & (\%)                   & $\left(^\circ\right)$ & $\left(^\circ\right)$\\
\hline                                    
CMa R1  & 2013 Jan 24 & $2.34$ & $0.94$ & $85.52$  & $56.0$\\
No.24   & 2013 Jan 25 & $1.73$ & $1.11$ & $96.78$  & $56.0$\\
        & 2013 Jan 26 & $2.42$ & $1.11$ & $78.24$  & $56.0$\\

\hline                                  
\end{tabular}
\end{table}
For ISAAC/VLT, we find a dependency between the instrumental polarization and the airmass of the object. This
effect varies the observed Stokes parameters, Q and U, significantly on scales of $10~\%$ and  is taken into account in the data analysis (see
Appendix~\ref{app:airmass} for details). For the SOFI/NTT data set, we applied a correction technique based on the correction of the ISAAC/VLT data. As a
successful test, the observational results of the polarized standard star of the SOFI/NTT observations after bias correction correspond well to the
literature values (see Tables~\ref{table:standards_summary},
\ref{table:standards_observations}).

\section{Polarization maps}\label{sec:polmap}

\begin{figure*}
\centering
\includegraphics[width=0.9\hsize]{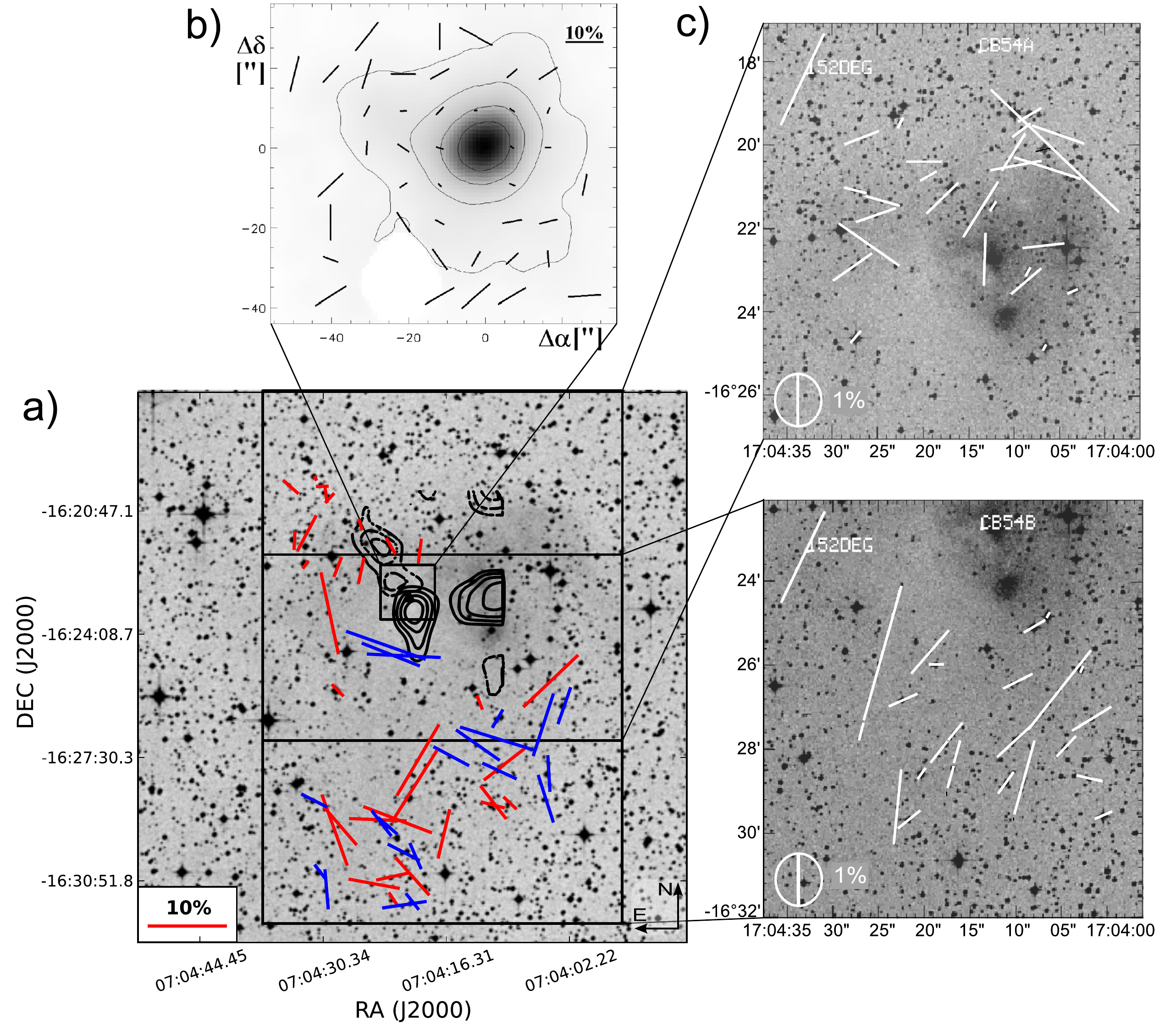}
\caption{CB54 - {\it a)} DSS-map with overlaid near-IR polarization vectors observed with ISAAC/VLT. The length of the vectors shows the degree of
polarization, and the direction gives the position angle. Only vectors with $P/\sigma_{\rm{P}} > 3 $ are plotted. Blue vectors have a better error statistic
than
the red vectors, since they are based on more than one set of observations.
The contour lines represent the $^{12}\text{CO}(1-0)$ spectral channel maps obtained with the Five College Radio Astronomy Observatory. The solid contour lines
give the blueshifted (south-western), and the dotted contour lines give the redshifted (north-eastern) outflow lobe. Contour levels are spaced at
$0.3~\text{K~km~s}^{-1}$ intervals of $1.7~\text{K~km~s}^{-1}$ (blue), and $0.15~\text{K~km~s}^{-1}$ intervals of $0.45~\text{K~km~s}^{-1}$ (red),
respectively \citep[from][beam size of $48''$; reproduced by permission of the AAS]{1994ApJS...92..145Y}. {\it b)} Intensity map overlaid with polarization
vectors obtained with JCMT/SCUBA in the sub-mm \citep[][reproduced by permission of the AAS]{2001ApJ...561..871H} {\it c)} Intensity map overlaid with
polarization vectors obtained with IMPOL/GIRT in the optical \citep[][]{2000A&AS..141..175S}.}
\label{fig:CB54_polmap}
\end{figure*}

In this section, we present the polarization maps in the sub-mm, in the near-IR, and in the optical of the three Bok globules B335, CB68, and CB54. For
comparing the sub-mm maps with the near-IR and optical maps, it has to be taken into account that the polarization arises from two different physical
processes: while thermal emission is observed in the sub-mm, dichroic absorption is observed in the near-IR and the optical. Thus, the polarization of a
dust grain observed in the sub-mm is oriented perpendicular to the polarization of the very same dust grain observed in the near-IR or optical.\\
The sub-mm maps of our three objects are each available in two versions, the originally published maps by
\cite{2003ApJ...592..233W}, \cite{2003ApJ...588..910V}, and \cite{2001ApJ...561..871H}, as well as rereduced maps by \cite{2009ApJS..182..143M}.
\cite{2009ApJS..182..143M} reduced all polarization observations again that have been executed with SCUBA/JCMT and cover a huge variety of objects, from
planets up to galaxies, in a single procedure. In doing so, \cite{2009ApJS..182..143M} applied very strict criteria for the decision of including or omitting
science data of an object and remark that data that was omitted by these generalized regulations may still be usable, and even part of publications, but would
require too much individual assessment to meet the requirements of their systematic rereduction. For this reason, we decided to use the originally reduced and
published sub-mm polarization maps in our
discussion (see, Fig.~\ref{fig:B335_polmap},~\ref{fig:CB68_polmap},~\ref{fig:CB54_polmap}).
\\
In the following, we discuss the polarization maps seperately for each observed wavelength, as well as in comparison to the maps of
the additional wavelength ranges.

\subsection{Bok globules with simple morphology: B335 and CB68}\label{subsec:B335CB68}
\begin{figure*}
\centering
\includegraphics[width=\hsize]{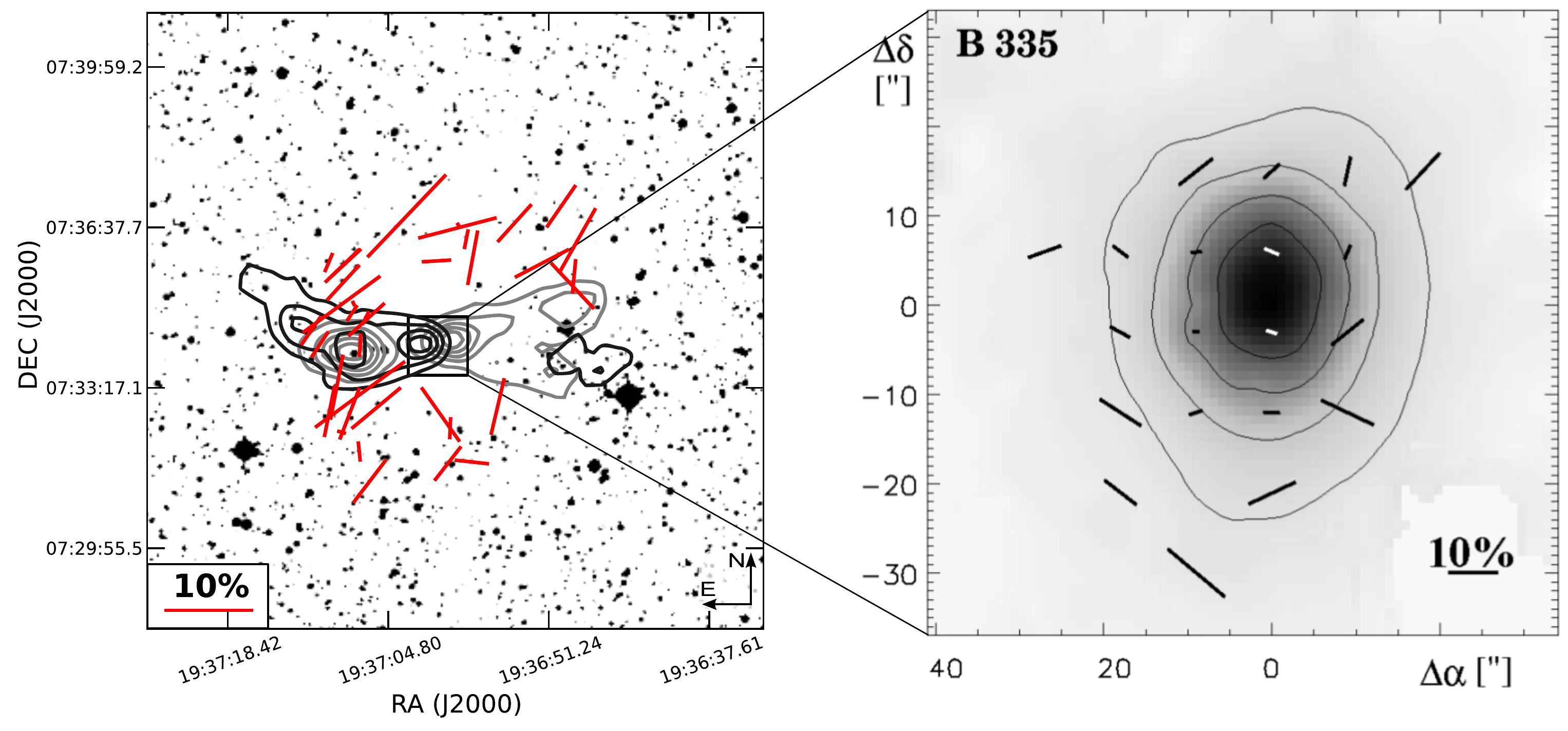}
\caption{B335 - {\it Left:} DSS-map with overlaid near-IR polarization vectors observed with ISAAC/VLT. The length of the vectors shows the degree of
polarization, and the direction gives the position angle. Only vectors with $P/\sigma_{\rm{P}}>3$ are plotted. The contour lines represent the
$^{12}\text{CO}(2-1)$
OTF channel maps obtained with the HHT on Mount Graham. The black (gray) contour lines give the blueshifted (redshifted) eastern (western) outflow lobe. The
contour levels are $\{1, 2, 3, 4 , 5, 6, 8, 10, 12, 14, 16\} \times 0.5~\text{K} - T_A^*$ \citep[from][beam size of $32''$; reproduced by permission of the
AAS]{2008ApJ...687..389S}. {\it Right:} Intensity map overlaid with polarization vectors obtained with SCUBA/JCMT in the sub-mm \citep[][reproduced by
permission of the AAS]{2003ApJ...592..233W}.}
\label{fig:B335_polmap}
\end{figure*}

\begin{figure*}
\centering\hspace*{6em}
\includegraphics[width=0.9\hsize]{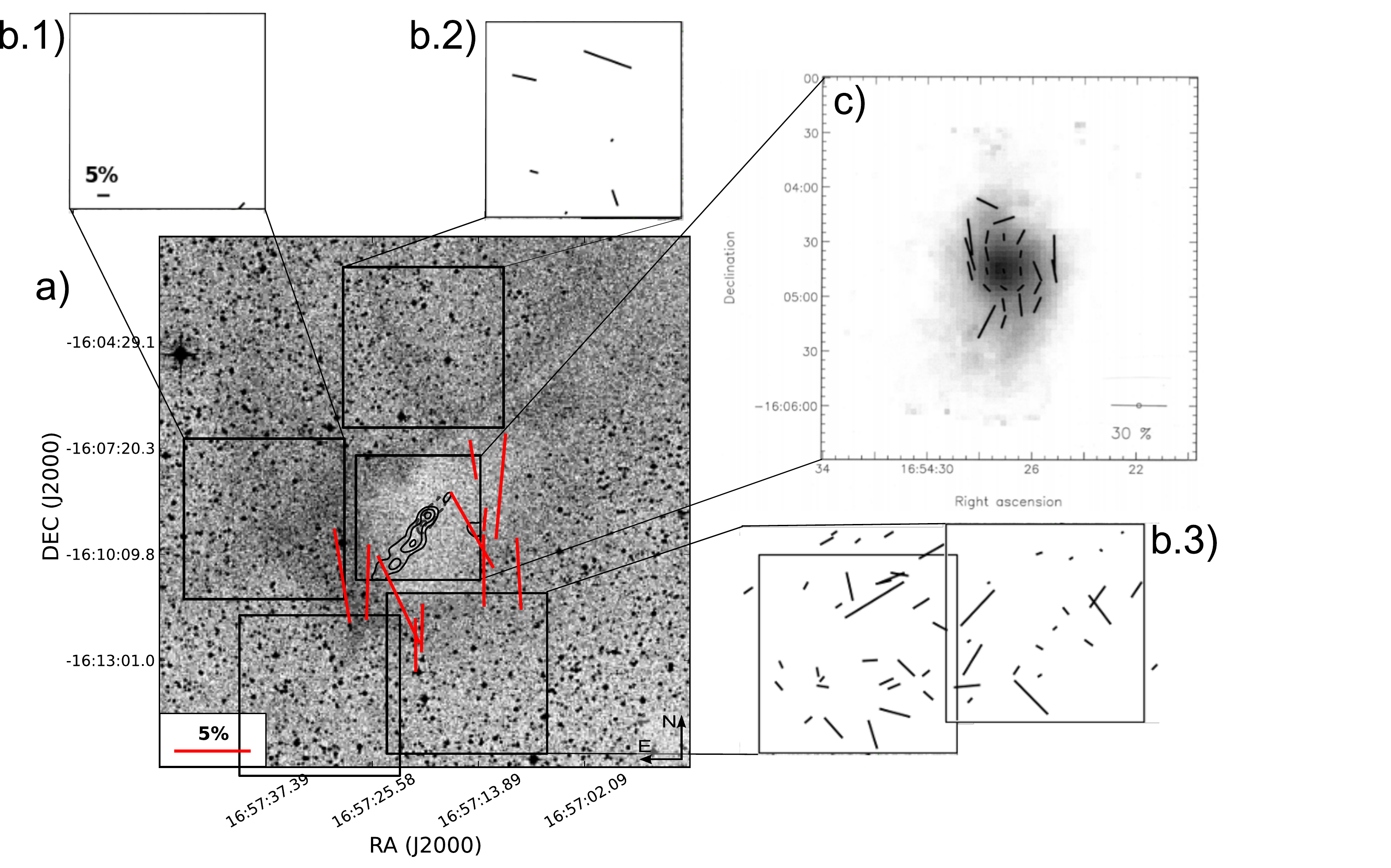}
\caption{CB68 - {\it a)} DSS-map with overlaid near-IR polarization vectors observed with ISAAC/VLT. The length of the vectors shows the degree of
polarization, and the direction gives the position angle. Only vectors with $P/\sigma_{\rm{P}} > 3 $ are plotted. The contour lines represent the
$^{12}\text{CO}(3-2)$
spectral channel maps obtained with the JCMT. The black contour lines give the blueshifted (south-eastern), the grey, dotted contour lines give the redshifted
(north-western) outflow lobe. Contour levels at $0.25, 0.50, 0.75, ...$~K \citep[from][HPBW of $14''$; reproduced by permission of the
AAS]{2000ApJ...542..352V}.
{\it b.1-3)} Same as {\it a)} but with polarization vectors obtained with IFOSC/IGO in the optical. {\it c)} Intensity map overlaid with polarization vectors
obtained with SCUBA/JCMT in the sub-mm \citep[][reproduced by permission of the AAS]{2003ApJ...588..910V}.} 
\label{fig:CB68_polmap}
\end{figure*}

\begin{figure*}
\centering
\includegraphics[width=0.9\hsize]{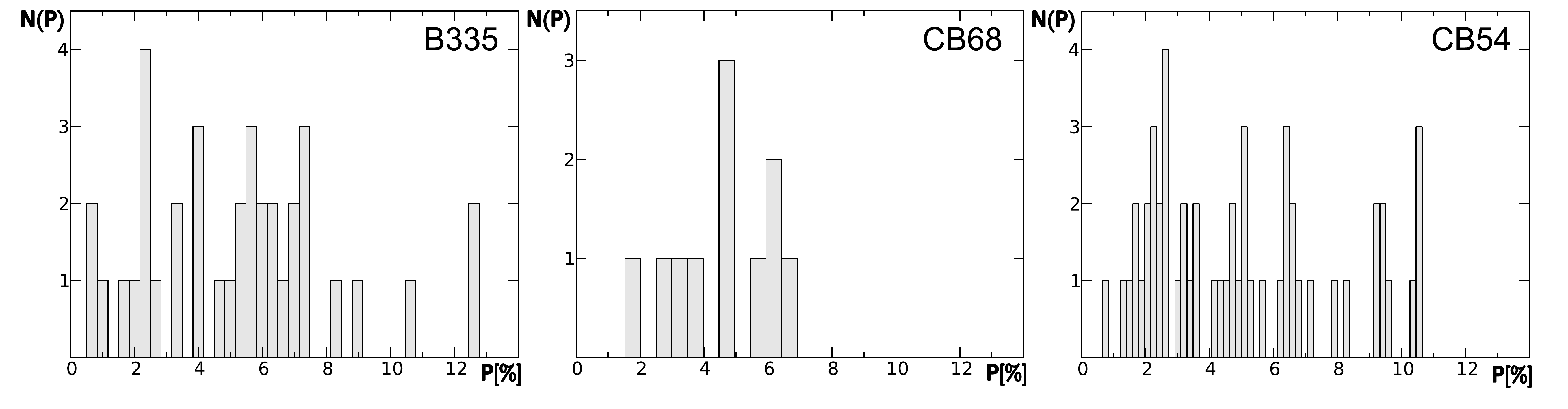}
\caption{Histograms showing the distribution of the degree of polarization, P for $P/\sigma_{\rm{P}} > 3 $, counts given by N(P), of B335, CB68, and CB54,
observed in
the near-IR.} 
\label{fig:deg_hists}
\end{figure*}

\begin{figure*}
\centering
\includegraphics[width=0.9\hsize]{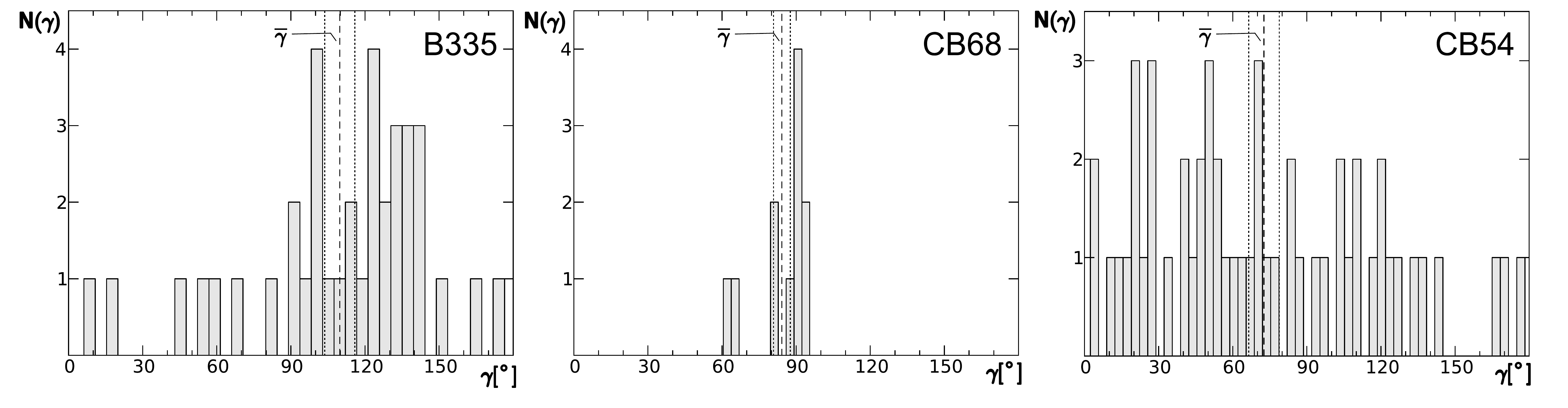}
\caption{Histograms showing the distribution of the polarization angle, $\gamma$, counts given by N($\gamma$), of B335, CB68, and CB54, observed in the
near-IR. The dashed lines represent the mean polarization angles, $\overline{\gamma}$, with the corresponding $1\sigma$ levels (dotted lines).} 
\label{fig:ang_hists}
\end{figure*}

The polarization maps of the Bok globules B335 and CB68 are shown in Figures~\ref{fig:B335_polmap} and \ref{fig:CB68_polmap}. Both globules disclose a
significant polarization signal $\left(P\gtrsim1~\%,\ P/\sigma_{\rm{P}}>3\right)$ in their outer, less dense parts in the near-IR as well as in the optical
for CB68 (see Fig.~\ref{fig:deg_hists}). The near-IR polarization vectors appear ordered very well in the case of CB68
($\overline{\gamma}=84.21\degr~\pm~3.39\degr$, see Fig.~\ref{fig:ang_hists}) and largely well-ordered in the case of B335
($\overline{\gamma}=103.94\degr~\pm~5.01\degr$, see Figs.~\ref{fig:B335_polmap},~\ref{fig:ang_hists}). According to ESO regulations, there are no observations
of polarized standard stars for the near-IR data observed with ISAAC/VLT. Thus, no independent quantitative measure of the reliability of the observed
polarization angles exists.\\
In both globules, a clear spatial gap remains between the polarization vectors in the sub-mm and in the near-IR. The gap width
in B335 amounts to about $70'' (\approx 7\times10^3~\text{AU})$ and to about $100'' (\approx 1.6\times10^4~\text{AU})$ in CB68. \citet{2013A&A...551A..98L}
observed the hydrogen column densities, $N_{\text{H}}$, of CB68 and B335. In the region traced by SCUBA/JCMT, they determine
$N_{\text{H}}\gtrsim10^{22}$cm$^{-2}$ and in the region traced by ISAAC/VLT and SOFI/NTT they find $N_{\text{H}}\lesssim10^{21}$cm$^{-2}$. This gives us a
benchmark for the feasibility to observe the morphology of the globules with the applied observing techniques and instrumentation and explains the
gap.\\
In the case of B335, the orientation of the near-IR polarization vectors in the eastern and southern part of the globule fits to the orientation of the sub-mm
polarization vectors very well. In the north-western part the polarization orientation shows local varieties in the sub-mm but not in the near-IR, even though
the near-IR polarization pattern appears slightly unordered (see, Fig.~\ref{fig:B335_polmap}).\\
In the case of CB68, there are near-IR polarization vectors with $P/\sigma_{\rm{P}} > 3 $ only in the western and in the south-eastern part.
\citet{2003ApJ...588..910V} observed a spiral polarization pattern in the sub-mm. One polarization 'spiral-arm' points toward
the southeastern globule part and thus, fits well to the near-IR polarization vectors. For the polarization vectors in the western part, this is not the case.
However, one has to keep in mind that there can also be an influence by that arm of the globule that stretches in the northwestern direction.
The optical polarization vectors of CB68 in the most outer parts of the globule are rather unordered. However, the polarization vectors in the northern part
(see Fig.~\ref{fig:CB68_polmap}~b.2) follow the smaller gas-dust branch that is perpendicular to the north-western arm of CB68. Additionally, in the southern
part (see Fig.~\ref{fig:CB68_polmap}~b.3, upper half of the left field) the optical polarization vectors follow the small extension of the globule, too.
There is a slight deviation of the orientation of the polarization vectors observed in the optical from those observed in the near-IR. The most straight
forward explanation is that the different vectors trace the magnetic field in different regions of the globule.


\subsection{Bok globule with more complex morphology: CB54}
The polarization map of the Bok globule CB54 is shown in Figure~\ref{fig:CB54_polmap}. We observe a significant polarization signal $\left(P\gtrsim2~\%,\
P/\sigma_{\rm{P}}>3\right)$ in the near-IR across the entire globule (see Fig.~\ref{fig:deg_hists}). As in the case of the globules CB68 and B335, we also
find a clear spatial gap with a width of about $1'$~(corresponding to about $7\times10^4~\text{AU})$ between the polarization vectors
observed in the sub-mm and in the near-IR in CB54.\\
Figure~\ref{fig:CB54_polmap} displays polarization vectors that cover the globule on scales from $10^3 - 10^4$~AU up to $10^5$~AU. Given the
complex structure of this Bok globule, it is not necessarily expected to find uniformly ordered polarization vectors across such large scales. However, on
smaller
scales ($\sim 10^4$~AU), by a factor of $\gtrsim2$ larger than the region traced by the sub-mm map, our near-IR observations and the optical
obvervations by
\citet{2000A&AS..141..175S} reveal ordered polarization vectors.\\
The southern field of the optical data (Fig.~\ref{fig:CB54_polmap}~c, lower field) overlaps with a part of our near-IR data.
The optical polarization degree is about one order of magnitude stronger than that in the near-IR. This factor is significantly
higher than the difference of about a factor of two which we expect from dust properties obtained from Mie theory \citep{1908AnP...330..377M}.
However, in terms of the orientation, the optical and near-IR vectors fit well to each other ($\Delta\gamma\lesssim10\degr$).\\
The region around the sub-mm map is of particular interest. In the sub-mm, \citet{2003ApJ...592..233W} observed randomly oriented
polarization vectors. However, our near-IR observations reveal ordered polarization vectors on larger scales around the sub-mm map (see
Fig.~\ref{fig:CB54_polmap}~a).

\section{Magnetic fields}\label{sec:Bfield}
In our analysis we assume that the magnetic field is oriented perpendicular to the measured polarization vectors in the sub-mm and parallel oriented
to the measured polarization vectors in the near-IR and in the optical. This widely applied concept is based on the finding that, independently of the alignment
mechanism, charged interstellar dust grains would have a substantial magnetic moment leading to a rapid precession of the grain angular momentum around the
magnetic field direction, implying a net alignment of the grains with the magnetic field \citep[see, e.g.,][]{1997ApJ...480..633D, 2007JQSRT.106..225L}.
However, one has to keep in mind that polarization observations strongly suffer from projectional effects along the line-of-sight (LOS). Thus, for a
comprehensive understanding of the magnetic field structure additional 3D radiative transfer modeling is essential but beyond the scope of this study.\\
In general, a high degree of polarization is connected to a magnetic field that is strong enough to align enough dust grains along the LOS to prevent the
annihilation of the polarized signal that is expected from randomly oriented grains along the LOS. In our observations of all three Bok
globules, B335, CB68, and CB54, we find strongly ordered magnetic fields. Additionally, our observations show well-ordered polarization vectors on
scales of $\lesssim 10^4$~AU for all three Bok globules. Thus our observations imply magnetic fields that are strongly ordered over large parts of
the globules. And finally, in the case of B335 and CB68, our near-IR and optical observations trace polarization patterns that fit to the polarization patterns
observed in the sub-mm by \cite{2003ApJ...592..233W} and \citet{2003ApJ...588..910V}, which implies strong and ordered magnetic fields on scales
from $10^2-10^3~\text{AU}$ to $10^4~\text{AU}$.\\
Besides this analysis of the magnetic field, \cite{1953ApJ...118..113C} give an estimate for the magnetic field strength in the
plane-of-sky (POS; CF~method):
\begin{equation}
 \left|\boldsymbol{B_{\rm{pos}}} \right| = \sqrt{\frac{4\pi}{3}~\varrho_{\rm{gas}}}~~\frac{v_{\rm{turb}}}{\sigma_{\gamma}},
\end{equation}
where $\varrho_{\rm{gas}}$ is the gas density in units of g~cm$^{-3}$, $v_{\rm{turb}}$ the rms turbulence velocity in units of cm~s$^{-1}$, and
$\sigma_{\gamma}$ the standard deviation of the polarization position angles in radians.
Here, it is assumed that the magnetic field is frozen in the cloud material. For a detailed discussion of this equation, we refer to
\cite{2012ARA&A..50...29C}, and for an extension of the mean magnetic
field strength, we refer to \cite{2004ApJ...616L.111H}.\\
For B335, we use the hydrogen number densities and the turbulence velocity from \cite{1987ApJ...313..320F}, for details see Table~\ref{table:CF}. To
account for helium and heavier elements, we derive the total gas density $\rho_{\rm{gas}}$ from
\begin{equation}
 \rho_{\rm{gas}} = 1.36~n_{\rm{H_2}}~M_{\rm{H_2}},
\end{equation}
where $M_{\rm{H_2}}=2.0158$~amu is the mass of a $\rm H_{\rm{2}}$ molecule. The corresponding values are listed in Table~\ref{table:CF}.\\
We can estimate the mean magnetic field strength in the region north of the center, northwest of the center, and east of the center of B335 (see
Table~\ref{table:CF}). We find similar magnetic field strengths in the different regions, $B_{\rm{B335,~near-IR}}~\approx~(12-40)~\mu$G, which are
less, by a factor of $3-10$, than the estimated field strength in the sub-mm, $B_{\rm{B335,~smm}}~\approx~130~\mu$G \citep[][]{2003ApJ...592..233W}.
Considering the inaccuracies in the estimations of the densities, the temperatures, and the gas velocities, we do not find any significant difference in
the magnetic field strengths in the inner and outer parts of B335.\\
For CB68 and CB54, information about the gas densities, temperatures, and turbulence velocities in the parts of the globules that we traced in
the near-IR and in the optical is not available. {\rm However,} in the sub-mm, \cite{2003ApJ...588..910V} and \cite{2001ApJ...561..871H} estimate the magnetic
field strengths of CB68 and CB54 to $B_{\rm{CB68,~smm}}~\approx~150~\mu$G and $B_{\rm{CB54,~smm}}~\approx~60~\mu$G.

\begin{table*}
\caption{ Gas densities, gas velocities, polarization, and magnetic field strengths of the outer parts of B335 traced in the near-IR.}           
\label{table:CF}      
\centering          
\begin{tabular}{l c c c c c c c c c }
\hline\hline   
Region & $n_{\rm{H_2}}$ & $\rho_{\rm{gas}}$ & $T_{\rm{kin}}$ & $\Delta v$ & $v_{\rm{turb}}$ & $N_{\rm{vec}}$ & $\bar{\gamma}$ & $\sigma_{\gamma}$ &
$B$ \\
& (cm$^{-3}$)& (g~cm$^{-3}$) & (K) & (km~s$^{-1}$) & (km~s$^{-1}$) & & (deg) &  (deg) & ($\mu$G) \\
\hline                    
center-north & $1.3\times10^3$ $^{\textrm{(a)}}$ & $5.92\times10^{-21}$ & $10$ $^{\textrm{(a)}}$ & $0.82\pm0.01$ $^{\textrm{(a)}}$ & $0.74$ & $5$ &
$135.42^{\circ}$ & $31.59^{\circ}$ & $21.14$\\
center-east  & $1.3\times10^3$ $^{\textrm{(a)}}$ & $5.92\times10^{-21}$ & $10$ $^{\textrm{(a)}}$ &  $0.82\pm0.01$ $^{\textrm{(a)}}$ &  $0.74$ & $9$ &
$122.89^{\circ}$ & $16.72^{\circ}$ & $39.93$\\
northwest	 & $1.8\times10^3$ $^{\textrm{(a)}}$ & $8.19\times10^{-21}$ & $10$ $^{\textrm{(a)}}$ &  $0.73\pm0.02$ $^{\textrm{(a)}}$ &  $0.62$ & $6$ &
$111.84^{\circ}$ & $33.61^{\circ}$ & $19.58$\\
			 & $6.5\times10^2$ $^{\textrm{(a)}}$ & $2.96\times10^{-21}$ & $20$ $^{\textrm{(a)}}$ &  $0.73\pm0.02$ $^{\textrm{(a)}}$ &  $0.62$ & $6$ &
$111.84^{\circ}$ & $33.61^{\circ}$ & $11.69$\\\hline                  
\end{tabular}
\tablebib{(a)~\citet{1987ApJ...313..320F}.}
\end{table*}

\section{Correlation between magnetic field structure and the CO~outflow of the Bok globules}\label{sec:CO}
Magnetic fields are believed not only to influence the collapse of Bok globules but also the formation of circumstellar disks and outflows. Many
observations show a preferential alignment of the outflow axis along the cloud-scale magnetic field \citep[e.g.,][]{1984MNRAS.210..425C, 1986AJ.....92..633V,
2003AJ....125.1418J}, but misaligned orientations of the outflow axes and magnetic field direction have been reported as
well \citep[e.g.,][]{2013ApJ...768..159H,
2003ApJ...592..233W}. Based on MHD~simulations, \citet{2006ApJ...637L.105M} find that the degree of alignment between outflow and magnetic field depends on the
magnetic field strength: the stronger the magnetic field, the better the alignment. However, one has to keep in mind that these simulations consider only
the most inner regions of globules with outflows on scales of $150-200$~AU. Thus, it is uncertain whether the findings of \citet{2006ApJ...637L.105M} can be
extended to the scales of our observations.\\
In the following, we examine the relative positions of the outflow axes and the magnetic field directions for our observations of the three Bok globules, B335,
CB68, and CB54, which all contain protostellar cores and drive collimated CO~outflows \citep[][]{1993ApJ...414L..29C, 2000ApJ...542..352V, 1994ApJS...92..145Y}.
\\
In discussing the relative orientation between the CO outflow and the magnetic field structure, one has to consider that only
one component of the spatial orientation of the outflow $\left(v \sin i\right)$ is known from velocity measurements and that polarization vectors only trace
the magnetic field structure projected on the plane of sky.  In the sub-mm, \citet{2003ApJ...592..233W} and \citet{2003ApJ...588..910V} find the polarization
vectors aligned with the CO~outflows and discuss the relation to the magnetic field structure traced by the polarization maps. We extend this analysis to the
outer, less dense regions traced by our near-IR and optical observations. Consistent with the sub-mm studies, we assume that the magnetic field is oriented
parallel to the observed polarization vectors in the near-IR and the optical (for details see Section~\ref{sec:Bfield}).
\\
{\bf B335} (Fig.~\ref{fig:B335_polmap}) -- Our near-IR observations trace the outflow region in the eastern part of this globule. Here, the near-IR polarization
vectors are oriented by an angle $\overline{\gamma} = 115.06\degr \pm 6.32\degr$ with respect to the outflow axis
$\left(\text{P.A.}~\approx~0\degr\right)$. The sub-mm polarization vectors are aligned nearly parallel to the outflow axis
\citep[][]{2003ApJ...592..233W}. The
globule itself is clearly elongated in the north-south direction (see Fig.~\ref{fig:B335_polmap}). Therefore, the magnetic field orientation changes from the
inner $7\times10^3$~AU, where it is perpendicular to the outflow and parallel to the globule, to the outer $2\times10^4$~AU, where
it is inclined toward the outflow.\\
{\bf CB68} (Fig.~\ref{fig:CB68_polmap}) -- The observed extent of the CO~outflow is limited to the inner $1.6\times10^4$~AU, and thereby traced
only by the near-IR polarization vectors in the southeastern part of this globule. These polarization vectors are oriented roughly perpendicular
$\left(\overline{\gamma}=78.95\degr~\pm~6.45\degr\right)$ to the outflow axis $\left(\text{P.A.}=142\degr \text{ and } -38\degr\right)$
\citep[][]{2000ApJ...542..352V}. Relative to the pseudodisk $\left(\text{P.A.}\approx51\degr\pm1\degr\right)$, which is located in the center 
of this globule ($<10^3$~AU), \citet{2003ApJ...588..910V} also find twisted sub-mm polarization vectors that they interpret as influenced by the
outflow. However, on scales of about $2\times10^3$~AU, the magnetic field orientation is nearly perpendicular to the outflow axis, and thereby almost parallel
to the pseudodisk.\\
{\bf CB54} (Fig.~\ref{fig:CB54_polmap}) -- Our near-IR observations trace the redshifted CO~outflow very well. We find polarization vectors that are almost
perfectly aligned with the outflow axis. This corresponds well to the findings of \citet{2003ApJ...592..233W} in the sub-mm, where a mean polarization
orientation perpendicular to the outflow direction was found. The blueshifted outflow lobe is barely traced by our near-IR polarization vectors that are almost
perpendicular to the outflow lobe. However, one has to consider that the distance between the polarization vectors and the CO~contour lines is large, and
the CO~outflow shows a dip toward the vectors. The magnetic field direction is clearly parallel to the redshifted CO outflow lobe on scales of about
$5\times10^4$~AU.\\
Following the discussion by \citet{2006ApJ...637L.105M}, our observations indicate that the magnetic field of B335 is stronger relative to the outflow in
the very outer parts of the globule than in the inner core and the other way around in the case of CB68, while the magnetic field of CB54 dominates the outflow
along its complete extent. These findings cannot be verified by our analysis of the magnetic field strength in B335 by using the CF~method (for
details, see
Sec.~\ref{sec:Bfield}). However, one has to keep in mind that these interpretation may be highly influenced by projectional effects.

\section{Summary}\label{sec:summary}
For the first time, we have obtained multiwavelength polarization maps that cover the extent of Bok globules over a range of $10^2-10^5$~AU, covering
optically thin and optically thick regions. We observed the three globules B335, CB68, and CB54 in the near-IR and in the optical, and combined our observations
with archival sub-mm and optical data. The major results follow.
\begin{enumerate}
 \item The polarization degrees in the near-IR and in the optical amount to several percent $\left(2~\%\lesssim P\lesssim10~\%\right)$.
 \item In the case of B335 and CB68, two simple structured Bok globules, the orientation of the sub-mm polarization vectors in the dense inner part
$(10^2-10^3~\text{AU})$ of the globules continue to the outer, less dense globule parts $(10^4-10^5~\text{AU})$.
 \item In the case of CB54, a globule with large-scale turbulences, the orientation of the polarization vectors in the near-IR is well-ordered on scales of
$\sim10^4~$AU.
 \item In the case of B335, we found comparable magnetic field strengths in the globule parts traced by our near-IR observations and in the parts traced by
sub-mm observations, by using the CF method.
 \item We do not find a general correlation between the magnetic field structure and the CO~outflow of the Bok globule. In CB54 we find a magnetic field
orientation parallel to the CO~outflow, while B335 shows a change in the orientation  of the magnetic field toward the outflow axis from the inner core to the
outer regions. In CB68, we find a magnetic field orientation nearly perpendicular to the CO~outflow.
 \item The instrumental polarization of ISAAC/VLT depends significantly on the airmass of the observed object. For our unpolarized standard star, EGGR118, we
determined the deviation of the polarization degree, $\Delta P\approx1.4~\%$.
\end{enumerate}
The well-ordered polarization vectors indicate dominant magnetic fields from scales of $10^2-10^3$~AU to scales of $10^4-10^5$~AU. In the particular case of
CB54, the randomly oriented polarization pattern in the sub-mm can be explained by, e.g., a change in the orientation from the region south of the sub-mm
map
to the region
north of the sub-mm map (see Fig.~\ref{fig:CB54_polmap}).\\
A gap remains with a width of about $1'-2'$ between the sub-mm observations done with SCUBA/JCMT and the near-IR observations performed with
ISAAC/VLT and SOFI/NTT of B335, CB68, and CB54. \citet{2013A&A...551A..98L} derive typical hydrogen column densities for the region traced by the SCUBA/JCMT
observations in Bok globules of $N_{\text{H}}\gtrsim10^{22}$~cm$^{-2}$, as well as $N_{\text{H}}\lesssim10^{21}$~cm$^{-2}$, for the regions observed with
ISAAC/VLT and SOFI/NTT. To close the gap and, finally, spatially connect the magnetic field observations from the smallest to the largest scale, obvervations
that trace the region of $10^{21}$~cm$^{-2}\lesssim N_{\text{H}}\lesssim10^{22}$~cm$^{-2}$ need to be done.
Based on the flux level determined at the edges of the sub-mm map of B335 and CB54 observed with SCUBA/JCMT \citep{2003ApJ...592..233W, 2001ApJ...561..871H}, we
estimate that these observations can be performed in less than one hour with ALMA since cycle$~2$.
\\
The key to understanding the influence of magnetic fields on the low-mass star formation process is knowledge about the three-dimensional structures of
the magnetic field and the object itself. Our observations, as well as all observations, suffer from projectional effects along the line of sight. Thus, it is
necessary to do three-dimensional modeling of Bok globules, which includes a proper description of dust, dust grain alignment, gaseous outflows, and
polarimetric radiative transfer, in addition to polarimetric observations. However, this type of analysis is beyond the scope of this study.

\begin{appendix}
\section{Airmass-dependent instrumental polarization}\label{app:airmass}
\begin{figure}
\centering
\includegraphics[width=0.9\hsize]{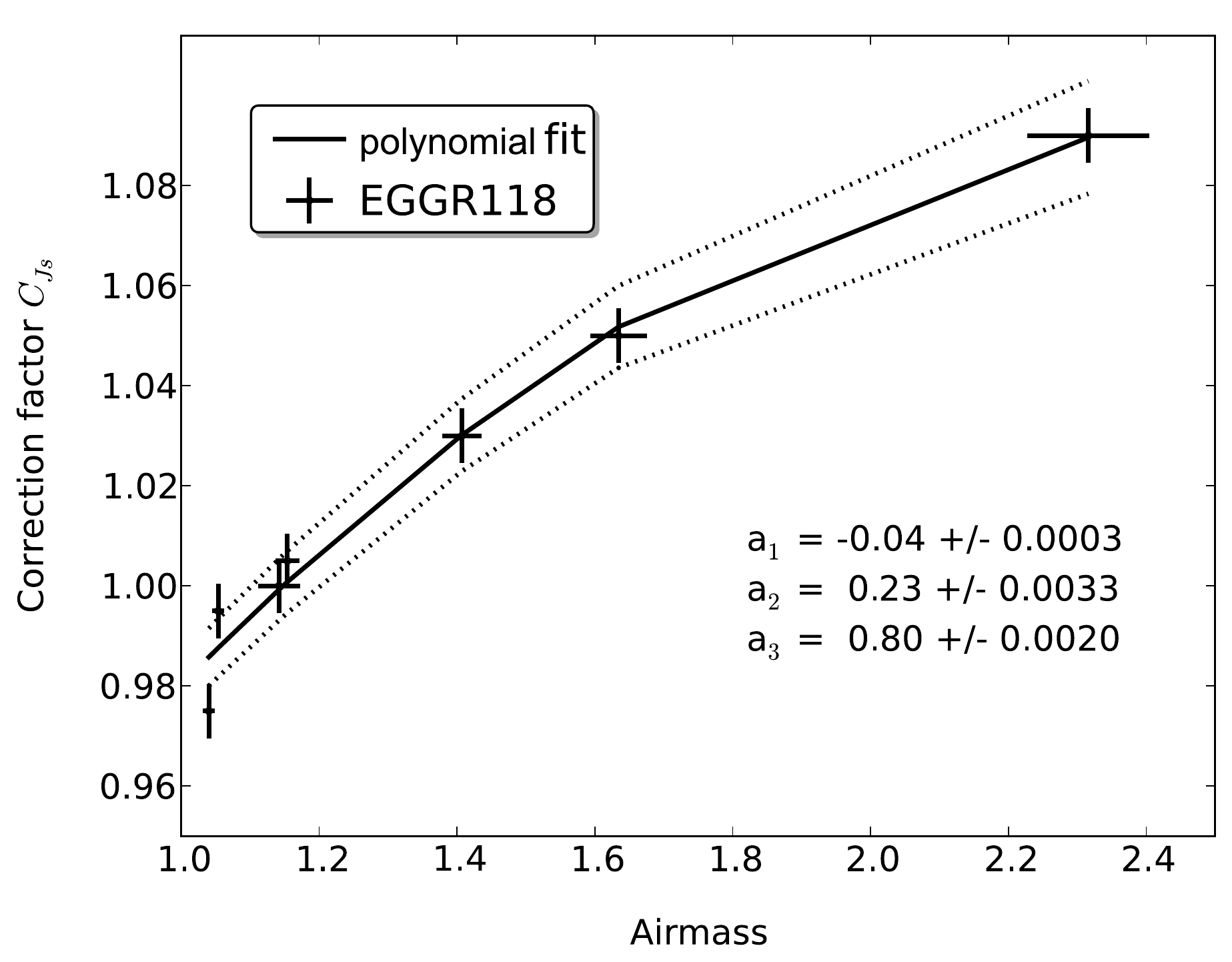}
\caption{The plot shows the determined correction factors, $C_{\rm{Js}}$, of EGGR118 over the corresponding airmasses and the fit to the data points, a
polynomial of
the 2nd degree. The coefficients of the polynomial are given by the a$_{\rm{i}}$, beginning with the leading coefficient.}
\label{appfig:calibration}
\end{figure}

\begin{figure}
\centering
\includegraphics[width=0.9\hsize]{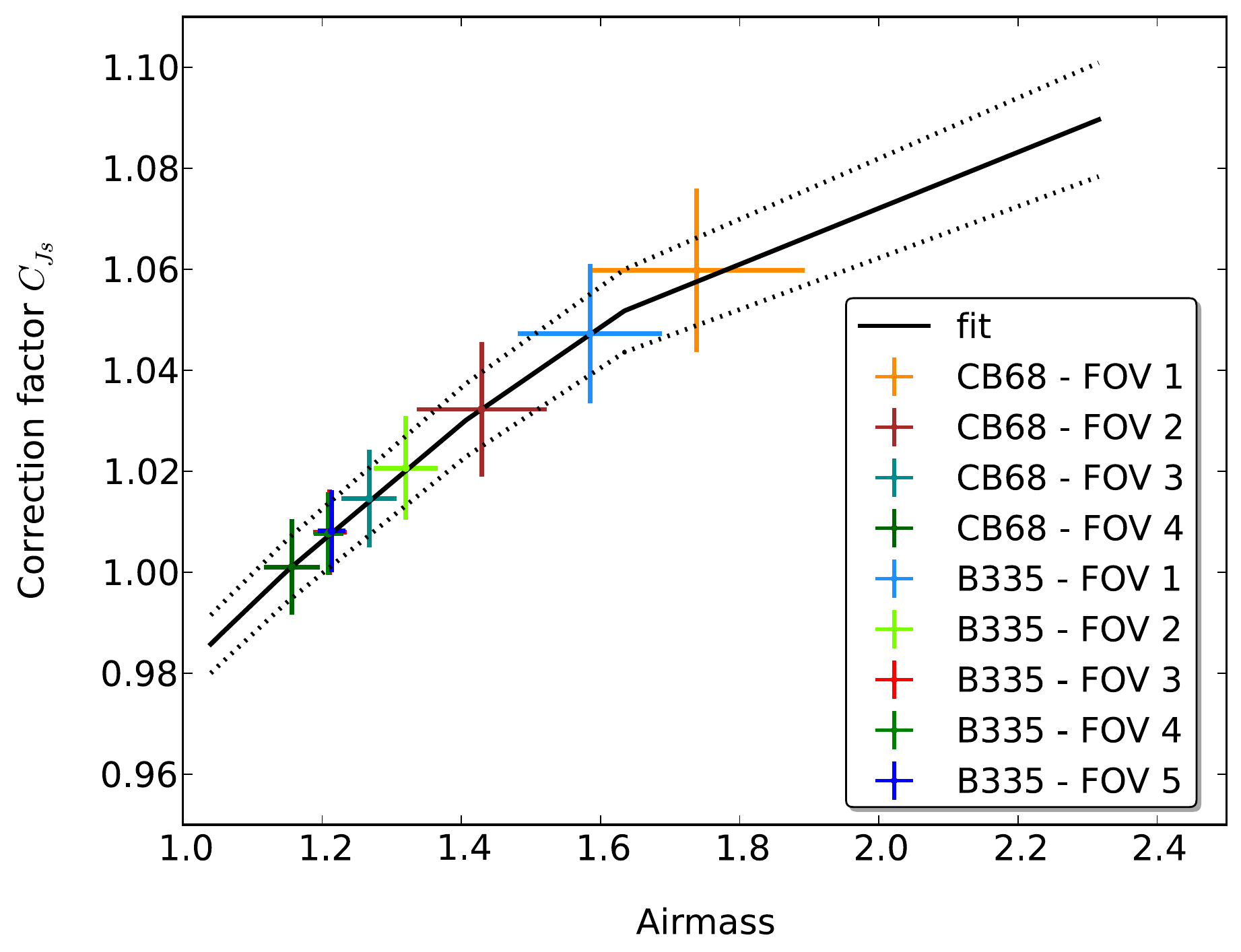}
\caption{We determine the corresponding correction factors, $C_{\rm{Js}}$, of the globules airmass intervals by using the polynomial fit (see
Fig.~\ref{appfig:calibration}). The errorbars in airmass direction give the airmass interval of the observation of each field. The errorbars in
$C_{\rm{Js}}$-direction give the error of the estimated correction factor of each data point, including photometric and fitting errors.}
\label{appfig:fit}
\end{figure}

Our ISAAC/VLT observations were performed in service mode between March 27 and May 27, 2012. Each night, we made two observations of unpolarized
standard stars, GJ1178 and EGGR118. When its right ascension meant that GJ1178 was no longer observable, two observations of EGGR118 were made in the same
night. Thus, we have observations of EGGR118 for the complete observing interval of two months, always performed with the very same procedure.\\
Every instrument or telescope may influence the polarization state of the incoming light, through its instrumental polarization. Therefore, the instrumental
polarization needs to be considered during the data reduction process. The assumption here is that the instrumental polarization is mainly created by a
non-ideal transmission ratio of the Wollaston prism. The ideal transmission ratio would be a 50~:~50 ratio of the intensities of the upper, $i_{\rm{u}}$, and
lower, $i_{\rm{l}}$, beams that are created by the Wollaston prism. To correct the signal for deviations in the ideal transmission ratio, a wavelength-dependent
correction factor, $C_{\rm{\lambda}}$, is applied \citep{HowToPol}:
\begin{eqnarray}
 i_{\rm{u},2} &=& i_{\rm{u},1} * C_{\rm{\lambda}}\\
 i_{\rm{l},2} &=& i_{\rm{l},1}~\mathtt{.}
\end{eqnarray}
To determine the instrumental polarization, $C_{\rm{\lambda}}$ has to be adjusted in such a way that the measured polarization of the unpolarized standard star
is minimal.\\
We determine an instrumental polarization that shows a strong dependence on the airmass of the object. In Fig.~\ref{appfig:calibration} we plot the
correction factor, $C_{\rm{Js}}$, over the airmass of EGGR118, observed in the Js~band throughout two months. The correction factor varies about $10~\%$ between
the beginning of the observations (airmass~$\approx 1$) to the end of the observations (airmass $\approx 2.4$). Since the Stokes parameters, Q and U, are
linearly dependent on $C_{\rm{\lambda}}$, the instrumental polarization itself changes significantly with the airmass of the observed object. For EGGR118, we
find $\Delta P\approx1.4~\%$.\\
To correct this effect, we fit a second-degree polynomial to the values of $C_{\rm{Js}}$ that we determine for the different airmasses at which we observed
EGGR118 (see Fig.~\ref{appfig:calibration}). Then, we take the airmass intervals of the Bok globule observations of each field and use the polynomial fit to
determine their corresponding correction factors (see Fig.~\ref{appfig:fit}).\\
The airmass dependency of the instrumental polarization clearly shows the invalidity of the assumption that the non-ideal Wollaston prism is the only
source of the instrumental polarization that has to be considered. Since ISAAC/VLT does not contain any components that move after the basic setup of the
instrument is completed, this dependency indicates significant influence by either the telescope itself or the atmosphere. Additional sources of the
instrumental polarization can be the deformation of the mirrors when the telescopes points to different airmasses or rather zenith angles, causing artifical
polarization in the moment when the beam is reflected.

\end{appendix}

\begin{acknowledgements}
Gesa Bertrang gratefully acknowledges financial support by the DFG under contract WO857/11-1 within the frame of the DFG Priority Program 1573: The Physics of
the Interstellar Medium.
\end{acknowledgements}

\bibliographystyle{aa}
\bibliography{lit}

\end{document}